\documentclass{article}

\usepackage{arxiv}

\usepackage[utf8]{inputenc} 
\usepackage[T1]{fontenc}    
\usepackage{hyperref}       
\usepackage{url}            
\usepackage{booktabs}       
\usepackage{amsmath}
\usepackage{amsfonts}       
\usepackage{nicefrac}       
\usepackage{microtype}      
\usepackage{cleveref}       
\usepackage{lipsum}         
\usepackage{graphicx}
\usepackage{natbib}
\usepackage{doi}
\usepackage{threeparttable}
\crefname{figure}{Figure}{Figures}
\crefname{table}{Table}{Tables}

\title{Stochastic Simulated Quantum Annealing for Fast Solution of Combinatorial Optimization Problems}

\author{
	Naoya Onizawa \\
	Research Institute of Electrical Communication\\
	Tohoku University\\
    Sendai, Japan 980-8577 \\
	\texttt{naoya.onizawa.a7@tohoku.ac.jp} \\
	\And
    Ryoma Sasaki \\
	Research Institute of Electrical Communication\\
	Tohoku University\\
	Sendai, Japan 980-8577 \\
	\texttt{ryoma.sasaki.p6@tohoku.ac.jp} \\
	\And
	Duckgyu Shin \\
	Research Institute of Electrical Communication\\
	Tohoku University\\
	Sendai, Japan 980-8577 \\
	\texttt{duckgyu.shin.p4@dc.tohoku.ac.jp} \\
	\And
	Warren J. Gross \\
	Department of Electrical and Computer Engineering\\
	McGill University\\
	Montreal, QC, Canada H3A 0E9 \\
	\texttt{warren.gross@mcgill.ca} \\
	\And
	Takahiro Hanyu \\
	Research Institute of Electrical Communication\\
   Tohoku University\\
   Sendai, Japan 980-8577 \\
  \texttt{takahiro.hanyu.c4@tohoku.ac.jp} \\ 
}

\renewcommand{\shorttitle}{Stochastic Quantum Monte Carlo Algorithm}

\hypersetup{
pdftitle={Stochastic Simulated Quantum Annealing  for Fast Solving Combinatorial Optimization Problems},
pdfsubject={SSQA},
pdfauthor={Naoya ~Onizawa},
pdfkeywords={Combinatorial optimization,stochastic computing, quantum monte carlo},
}

\begin{document}
\maketitle

\begin{abstract}

	In this paper, we introduce stochastic simulated quantum annealing (SSQA) for fast solution of combinatorial optimization problems.
	SSQA is designed based on stochastic computing and quantum Monte Carlo, which can simulate quantum annealing (QA) by using multiple replicas of spins (probabilistic bits) in classical computing.
	The use of stochastic computing leads to an efficient parallel spin-state update algorithm, enabling quick search for a solution around the global minimum energy.
	Therefore, SSQA realizes quantum-like annealing for large-scale problems and can handle fully connected models in combinatorial optimization, unlike QA.
	The proposed method is evaluated in MATLAB on graph isomorphism problems, which are typical combinatorial optimization problems.
	The proposed method achieves a convergence speed an order of magnitude faster than a conventional stochastic simulated annealing method.
	Additionally, it can handle a 100-times larger problem size compared to QA and a 25-times larger problem size compared to a traditional SA method, respectively, for similar convergence probabilities.
\end{abstract}



\section{Introduction}
\label{sec:introduction}

Combinatorial optimization is the process of searching for the optimal solution of an objective function in real-world applications, such as scheduling and machine learning \cite{QA_review}.
Simulated annealing (SA) \cite{SA1,SA2} is a potential approach for tackling NP-hard combinatorial optimization problems \cite{NP-hard}, where exact algorithms are ineffective.
Various SA methods have been developed to solve combinatorial optimization problems represented using an Ising model \cite{SA_max-cut,Ising_HW1,Ising_HW2}.
In addition to the classical approach, quantum annealing (QA) \cite{QA} is an alternative solution implemented using quantum devices, such as D-Wave quantum annealers \cite{D-Wave5000}.
QA is expected to solve large-scale combinatorial optimization problems faster than conventional SA methods; however, it currently faces limitations in handling small-size problems due to quantum device performance constraints \cite{QA_Google}.

Recently, a stochastic-computing-based SA (SSA) was proposed, which achieved orders of magnitude faster annealing compared to conventional SA and QA methods \cite{SSA}.
SSA employs a parallel form of simulated annealing by approximating probabilistic bits (p-bits) \cite{IL} using stochastic computing \cite{stochastic_first,stochastic}.
The combination of p-bits and stochastic computing enables rapid convergence to the global minimum of the objective functions.
The effectiveness of SSA has been evaluated on combinatorial optimization problems involving a few hundred bits, while real-world applications may involve a few thousand bits.

In this paper, we present stochastic simulated quantum annealing (SSQA)
for efficiently solving large-scale combinatorial optimization problems.
SSQA is a simulated quantum annealing (SQA) approach based on quantum Monte Carlo (QMC), which emulates quantum behavior on classical computers.
QMC approximates quantum bits by using multiple replicas of classical bits (spins) through the Trotter-Suzuki decomposition \cite{Suzuki, QMC}.
SQA is expected to outperform conventional SA in large-scale combinatorial optimization problems \cite{QA_Google}, and unlike QA, it can handle fully connected Ising models for combinatorial optimization problems \cite{Ising}.
In the proposed SSQA, we employ an efficient spin-state update algorithm using stochastic computing, similar to SSA \cite{SSA}, enabling quick exploration of the solution space around the global minimum energy.
The proposed algorithm is simulated using MATLAB on graph isomorphism (GI) problems with up to 2,500 spins, where GI is a typical combinatorial optimization problem represented by a fully connected Ising model.
The simulation results demonstrate that the proposed method achieves an order of magnitude reduction in time-to-solution (TTS) compared to and SSA and a few orders of magnitude reduction compared to the traditional SA.
Furthermore, when compared to experimental results of QA on the 504-qubit D-Wave Two machine \cite{QA_GI}, SSQA can handle problems approximately two orders of magnitude larger.

The rest of the paper is structured as follows.
\cref{sec:prel} provides a review of conventional SA methods and Ising models for combinatorial optimization problems.
\cref{sec:QMC} introduces the proposed SSQA algorithm based on stochastic computing.
\cref{sec:eval} compares the proposed algorithm with the traditional SA, SSA and QA methods.
Finally, \cref{sec:conc} concludes the paper.

\section{Preliminaries}
\label{sec:prel}

\subsection{Simulated annealing (SA) and Ising model}

\begin{figure}[t]
	\centering
	\includegraphics[width=1.0\linewidth]{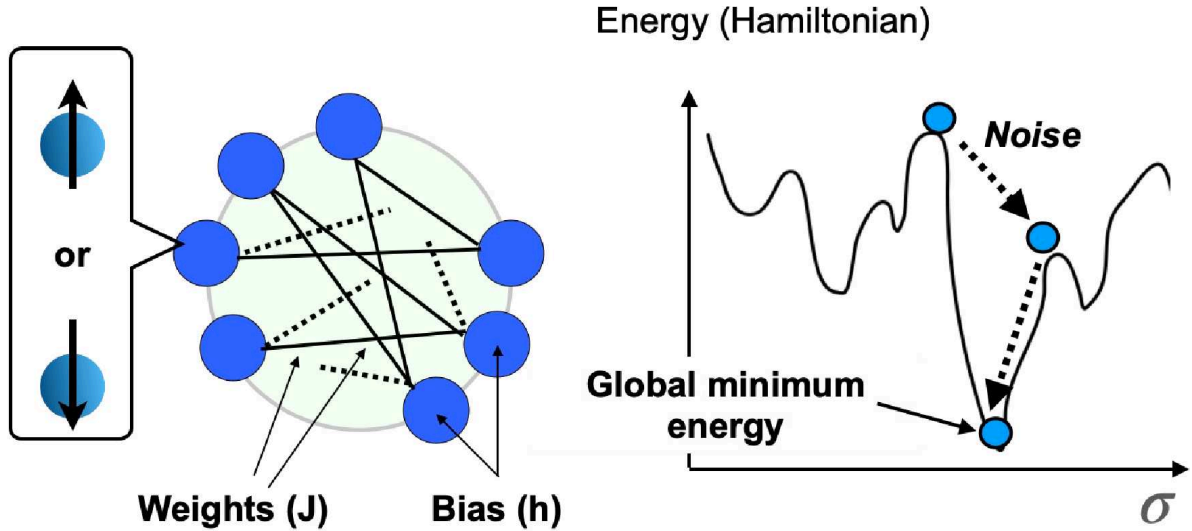}
	\caption{Simulated annealing (SA) based on a spin network that consists of  spins, spin biases, and spin weights. Spin states can be flipped between '+1' and '-1' in an attempt to reach the global minimum energy of the Hamiltonian.}
	\label{fig:SA}
\end{figure}

\cref{fig:SA} illustrates a spin network-based simulated annealing (SA) algorithm.
The spin network is constructed using spins, spin biases ($h$), and spin weights ($J$) connecting the spins.
The spin states ($\sigma$) can take two values: '-1' and '+1'.
The spin network is based on the Ising model \cite{Ising}, which represents a Hamiltonian (energy function) as follows:
\begin{equation}
	H(\sigma) = - \sum_i h_i\sigma_i - \sum_{i < j} J_{ij}\sigma_i\sigma_j.
	\label{eqn:Ising_SA}
\end{equation}
where $i$ and $j$ ($1 \leq i, j \leq N$) are indices of spins, and  $N$ is the number of spins.

SA is commonly applied to solve various combinatorial optimization problems, including GI, traveling salesman, and maximum cut problems \cite{QA_GI, max-cut, TSPLIB}.
These problems can be represented by the coefficients $h$ and $J$ in  \cref{eqn:Ising_SA}.
During the annealing process, spin states can be flipped between '+1' and '-1' in an attempt to reach the global minimum of the Hamiltonian.
Several SA methods have been proposed, such as serial updating \cite{SA_max-cut}, parallel updating \cite{Ising_HW1}, and parallel tempering \cite{Ising_HW2}, to enhance the efficiency of the annealing process.

\subsection{Stochastic-computing-based simulated annealing (SSA)}

Recently, a p-bit-based simulated annealing (pSA) approach was introduced in \cite{pbits_general}.
A p-bit is a probabilistic bit that can be in one of two spin states, '+1' and '-1'.
It has been proposed for invertible logic, which is an unconventional computing technique \cite{IL, CIL, CIL_training}.
pSA, implemented on an underlying Boltzmann machine \cite{Boltzmann1984}, enables parallel updating of spins for faster simulated annealing. 
However, it suffers from slow convergence to the global minimum energy.

To address this issue, the SSA method was proposed in \cite{SSA}.
SSA utilizes p-bits that are approximated using integral stochastic computing \cite{SDNN}.
It is  worth noting that integral stochastic computing is an extended version of stochastic computing \cite{stochastic_first, stochastic} and offers area-efficient hardware implementation \cite{Sldpc1, Simage, SIIR}.
The approximation of p-bits in SSA leads to a faster simulated annealing process compared to pSA.
In SSA, each spin state is updated as follows:
\begin{subequations}
	\begin{equation}
		I_i(t+1) = h_i+\sum_j J_{ij}\cdot \sigma_j(t) + n_{rnd}\cdot r_i(t),
		\label{eqn:I_SC-SA}
	\end{equation}
	\begin{equation}
		I{s}_i(t+1)=
		\begin{cases}
			I_0(t)-\alpha, \text{if} \ I{s}_i(t) + I_i(t+1) \geq I_0(t) \\
			-I_0(t), \text{else if} \ I{s}_i(t) + I_i(t+1) < -I_0(t) \\
			I{s}_i(t) + I_i(t+1),  \text{otherwise}
		\end{cases}
		\label{eqn:updown_SC-SA}
	\end{equation}
	\begin{equation}
		\sigma_i(t+1)=
		\begin{cases}
			1,& \text{if} \ I{s}_i(t+1) \geq 0 \\
			-1, & \text{otherwise},
		\end{cases}
		\label{eqn:m_SC-SA}
	\end{equation}
	\label{eqn:SC-SA}
\end{subequations}
where $\sigma_i(t) \in \{-1,1\}$ and $\sigma_i(t+1) \in \{-1,1\}$ represent the binary input and output spin states, respectively.
$I_0$ is the pseudo inverse temperature, $I_i(t+1)$ and $Is_i(t+1)$ are real-valued internal signals, and $n_{rnd}$ is the noise magnitude of a random signal $r_i(t) \in \{-1,1\}$.
$\alpha$ is the minimum resolution of data representation.
If only integer values are used in \cref{eqn:SC-SA}, $\alpha$ can be 1 \cite{SSA}.
Equations (\ref{eqn:updown_SC-SA}) and (\ref{eqn:m_SC-SA}) provide approximations of the $\tanh$ function for p-bits using integral stochastic computing.
During the annealing process in SSA, $I_0$ gradually increases while updating all spin states.
Since SSA is designed based on stochastic computing, it can be implemented in both software and hardware \cite{JETCAS_SSA}.

\begin{table*}[t]
	\caption{Summary of annealing methods for solving combinatorial optimization problems, including SA \cite{SA1}, QA \cite{QA}, and SSA \cite{SSA}.}
	\begin{threeparttable}
		\centering
		\begin{tabular}{c||c|c|c|c}
			\hline
			&  SA & QA  & SSA  & This work (SSQA)\\
			\hline
			\hline
			Fundamental  & Simulated annealing & Quantum annealing & Simulated annealing & Simulated  \\
			algorithm & & & & quantum annealing\\
			\hline
			Spin update & Serial & Parallel & Parallel & Parallel \\
			\hline
			Computational & Classical computing & Quantum computing & Classical computing & Classical computing\\
			model & & & (stochastic computing\tnote{a} ) & (stochastic computing\tnote{a} ) \\
			\hline
			Problem& No & Yes & No & No \\
			limitation & & & & \\
			\hline
		\end{tabular}
		\begin{tablenotes}
			\item[a] Stocahstic computing can be implemented using typical digital circuits.
		\end{tablenotes}
	\end{threeparttable}
	\label{tb:summary}
\end{table*}

\section{Stochastic Simulated Quantum Annealing (SSQA) Algorithm}
\label{sec:QMC}

\subsection{Hamiltonian}
\label{ssec:H_QMC}

\cref{tb:summary} provides a summary of annealing methods, including SA, QA, and SSA, all utilized for solving combinatorial optimization problems.
SA \cite{SA1} is a conventional simulated annealing method that updates a randomly selected spin in order to reach the global minimum energy.
QA \cite{QA} achieves parallel spin updates using quantum devices; however, its problem-solving capabilities are limited due to current device performance constraints.
SSA \cite{SSA} realizes fast annealing based on stochastic computing.
In this paper, we propose stochastic simulated quantum annealing (SSQA) as an extension of SSA.

Let us explain the Hamiltonian of SSQA.
SSQA is a type of simulated quantum annealing (SQA) approach based on quantum Monte Carlo (QMC), which can emulate the behavior of QA.
In contrast to SA, the Hamiltonian $H_{q}(\sigma)$ of QA is represented as follows:
\begin{equation}
	H_{q}(\sigma) = - \sum_i h_i\sigma_i^z - \sum_{i < j} J_{ij}\sigma_i^z\sigma_j^z - \Gamma_x \sum_i \sigma_i^x,
	\label{eqn:Ising_QA}
\end{equation}
where $- \sum_i h_i\sigma_i^z - \sum_{i < j} J_{ij}\sigma_i^z\sigma_j^z$ represents the problem Hamiltonian, and $\Gamma_x$ is a scheduling parameter for annealing.
Here, $\sigma_i^z$ and $\sigma_i^x$ are the Pauli matrices \cite{QA} that act on quantum devices used in quantum annealing machines, such as D-Wave \cite{D-Wave5000}.

\begin{figure}[t]
	\centering
	\includegraphics[width=0.6\linewidth]{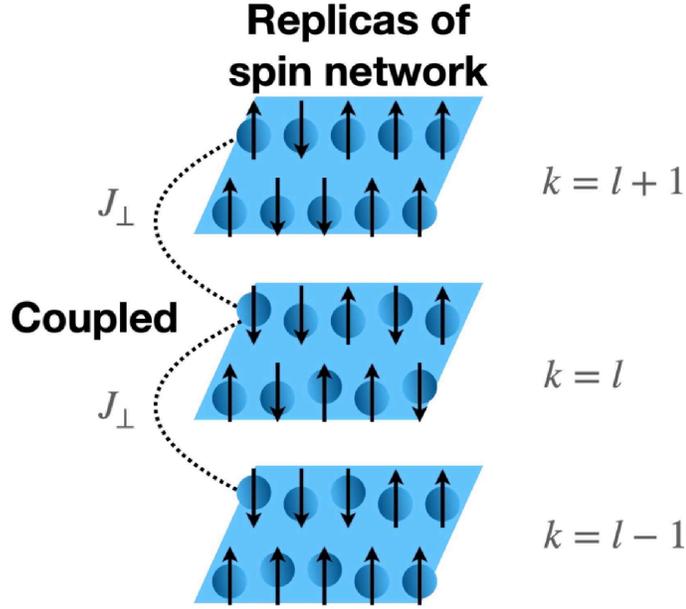}
	\caption{Quantum Monte Carlo (QMC) method using spin network replicas for SSQA. Each spin network is coupled with the upper and the lower spin networks with $J_{\perp}$.}
	\label{fig:QMC_model}
\end{figure}

The Trotter-Suzuki decomposition approximates the Hamiltonian in equation (\ref{eqn:Ising_QA}), enabling its representation using multiple replicas of spins on classical computers \cite{Suzuki, QMC}. 
This approximation is used by Quantum Monte Carlo (QMC) to simulate the process of QA.
The Hamiltonian of QMC, denoted $H_{c}(\sigma)$, can be expressed as follows \cite{pbits-QMC}:
\begin{equation}
	H_{c}(\sigma) = \sum_{k=1}^{R} \Bigl(H_{p}(\sigma)  
	-  J_{\perp}\sum_i \sigma_{i,k}\sigma_{i,k+1} \Bigr), 	
	\label{eqn:Ising_QMC}
\end{equation}
\begin{equation}
	H_p(\sigma) = - \sum_i h_i\sigma_{i,k} -  \sum_{i < j} J_{ij}\sigma_{i,k}\sigma_{j,k},
	\label{eqn:Ising_QMC2}
\end{equation}
where $H_p(\sigma)$ is the problem Hamiltonian, $\sigma_{i,k}$ represents the spin state of the $k$-th replica of the spin network ($1 \leq k \leq R$), $R$ is the number of replicas of spins used to represent the q-bit, and $J_{\perp}$ is a scheduling parameter corresponding to $\Gamma_x$.
The problem Hamiltonian, $H_p(\sigma)$, is the same as the one in equation (\ref{eqn:Ising_SA}) used in SA.

\cref{fig:QMC_model} illustrates a QMC method utilizing $R$ replicas of the spin network.
Each spin network represents the problem Hamiltonian and is loosely coupled to the upper and lower spin networks using $J_{\perp}$.
It is important to note that the top replica is coupled to the bottom one.

\begin{figure}[t]
	\centering
	\includegraphics[width=0.6\linewidth]{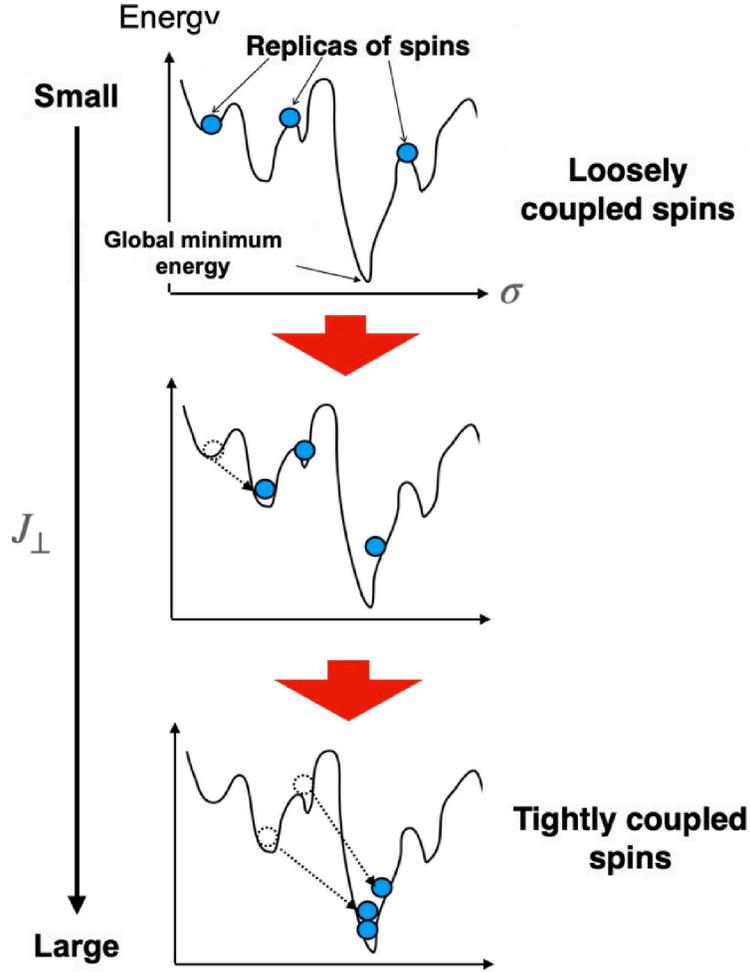}
	\caption{Concept of annealing process of SSQA. Each replica of the spin network searches for the global minimum energy with increasing $J_{\perp}$, which can reach the global minimum based on the coupled spins.}
	\label{fig:QMC_annealing}
\end{figure}

\subsection{Spin-update algorithm}
\label{ssec:SSQA}

\cref{fig:QMC_annealing} illustrates the concept of the annealing process in SSQA that performs based on QMC with equations (\ref{eqn:Ising_QMC}) and (\ref{eqn:Ising_QMC2}).
There are $R$ replicas of spins, which are coupled by $J_{\perp}$.
During the annealing process, spin states are randomly flipped in order to reach the global minimum energy of the problem Hamiltonian.
Each replica of the spin network independently searches for the global minimum energy with the effect of $J_{\perp}$.
When $J_{\perp}$ is small, the spin network is loosely coupled with the upper and lower replicas, allowing each replica to search for the global minimum energy of the problem Hamiltonian independently with minimal influence from the neighboring replicas.
On the other hand, when $J_{\perp}$ is large, the spin network becomes tightly coupled, enabling it to reach the global minimum energy by leveraging replicated spins with low energies.

SSQA is designed based on integral stochastic computing \cite{SDNN}.
The spin-update algorithm of SSQA is designed based on \cref{eqn:SC-SA} of SSA, as SSQA is an extension of SSA.
The update rule for the $i$-th spin state in the $k$-th replica is as follows:
\begin{subequations}
	\begin{eqnarray}
		I_{i,k}(t+1) &=& h_i+\sum_j J_{ij}\cdot \sigma_{j,k}(t)  + n_{rnd}\cdot r_i(t) \notag \\
		&+& J_{\perp}(t) \cdot \sigma_{i,k+1}(t-d),
		\label{eqn:I_SC-QMC}
	\end{eqnarray}
	\begin{equation}
		I{s}_{i,k}(t+1)=
		\begin{cases}
			I_0-\alpha, \text{if} \ I{s}_{i,k}(t) + I_{i,k}(t+1) \geq I_0 \\
			-I_0, \text{else if} \ I{s}_{i,k}(t) + I_{i,k}(t+1) < -I_0 \\
			I{s}_{i,k}(t) + I_{i,k}(t+1),  \text{otherwise}
		\end{cases}
		\label{eqn:updown_SC-QMC}
	\end{equation}
	\begin{equation}
		\sigma_{i,k}(t+1)=
		\begin{cases}
			1,& \text{if} \ I{s}_{i,k}(t+1) \geq 0 \\
			-1, & \text{otherwise}.
		\end{cases}
		\label{eqn:m_SC-QMC}
	\end{equation}
	\label{eqn:SC-QMC}
\end{subequations}
where $\sigma_{i,k}(t) \in \{-1,1\}$ and $\sigma_{i,k}(t+1) \in \{-1,1\}$ represent the binary input and output spin states, respectively.
Here, $\sigma_{i,k+1}(t-d)$ represents the $i$-th spin state in the $(k+1)$-th replica, and $d$ is the delay cycle for the coupled effect.
The coupled effect from the upper replica is represented by $J_{\perp}(t) \cdot \sigma_{i,k+1}(t-d)$.
$I_{i,k}(t+1)$ and $Is_{i,k}(t+1)$ are real-valued internal signals.

In SSQA, all the $(N \cdot R)$ spin states are updated in parallel.
Therefore, the computation cost of SSQA depends on the number of replicas $R$, which can influence the probability of convergence to the global minimum energy and the simulation time.

\section{Evaluation}
\label{sec:eval}

\subsection{Hamiltonian design for GI}

To evaluate the proposed SSQA algorithm, SSQA is simulated on the graph isomorphic (GI) problem, which is a typical combinatorial optimization problem \cite{QA_GI}.
A GI problem determines whether two graphs are isomorphic.
When solving a combinatorial optimization problem, it is first presented using a quadratic unconstrained binary optimization (QUBO) model.
The QUBO model is defined as follows:
\begin{equation}
	H(x) = \sum_{i,j} Q_{ij}x_ix_j
	\label{eqn:QUBO}
\end{equation}
where $Q_{ij}$ is an upper triangular matrix and $x_i \in \{0,1\}$ are binary variables.

\begin{figure}[t]
	\centering
	\includegraphics[width=0.6\linewidth]{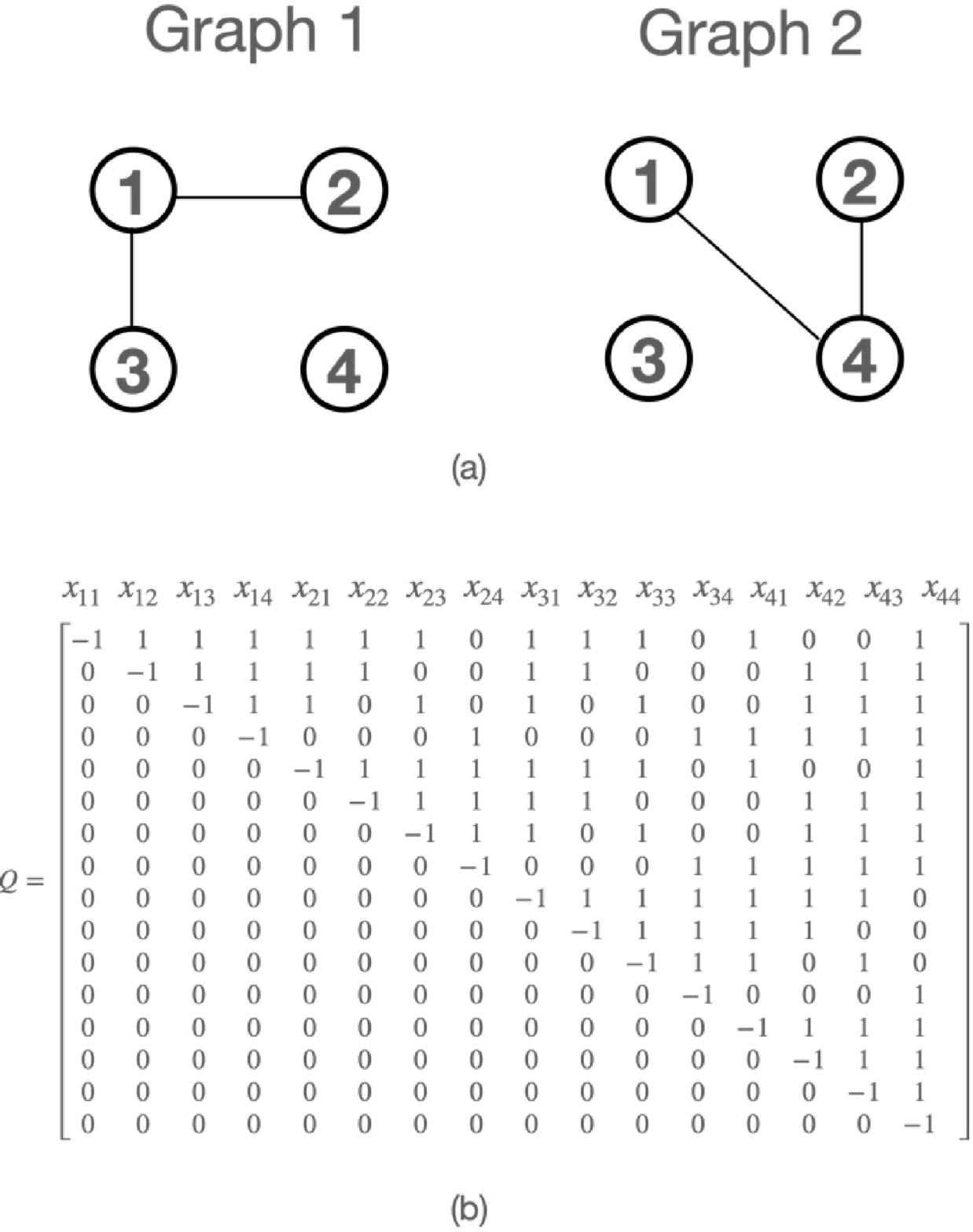}
	\caption{Example of a four-node graph isomorphism (GI) problem. (a) Two graphs are isomorphic and (b)  $Q$ coefficients of the QUBO model corresponding to the two four-node graphs.} 
	\label{fig:GI}
\end{figure}

\cref{fig:GI} (a) shows an example of the four-node GI problem.
In this example, Graph 1 and Graph 2 are isomorphic.
The QUBO model is obtained based on a vertex mapping penalty ($C_1$) and an edge inconsistency penalty ($C_2$) \cite{Ising}  as follows:
\begin{align}
	H(x) = C_1\sum_u(1-\sum_i x_{u,i})^2 + C_1\sum_i(1-\sum_u x_{u,i})^2 \nonumber \\
	+ C_2 \sum_{i,j \notin E_1, i\neq j}\sum_{u,v \in E_2} x_{u,i} x_{v,j} \nonumber \\
	+ C_2 \sum_{i,j \in E_1} \sum_{u,v \notin E_2, u\neq v} x_{u,i} x_{v,j},
	\label{eqn:H_GI}
\end{align}
where $x_{u,i}$ $\in \{0,1\}$ for every possible mapping of a vertex $u$ in Graph 2 to a vertex $i$ in Graph 1.
$Q$ is an $N \times N$ matrix, where $N$ is the square of the number of nodes in the GI problems.
Note that $N$ corresponds to the number of spins in the spin network.
The $Q$ coefficients of this example are shown in \cref{fig:GI} (b).

\subsection{Simulation setup}

In this simulation, firstly, a GI problem for each  graph size is generated according to \cite{QA_GI}. 
Let us explain the process for a five-node graph.
A graph with five nodes is randomly generated, where  a probability of connecting between two nodes is 50\%.
Hence, five edges exist on average as  there are 10 possible connections for the five-node graph.
The generated graph ('Graph 1') is copied to another graph ('Graph 2'), which then creates a QUBO model from equation (\ref{eqn:H_GI}).
Secondly, the QUBO model is converted into Hamiltonian coefficients $h$ and $J$ as described in equations (\ref{eqn:Ising_QMC}) and (\ref{eqn:Ising_QMC2}), with $\sigma_i = 2x_i-1$, $h_{i} = -\frac{1}{2}{Q_{ii}}-\frac{1}{4}\sum_{j \in \partial_i} Q_{ij}$, and $J_{ji} = J_{ij} = -\frac{1}{4}{\underset{i \neq j}{Q_{ij}}}$ (see details in \cite{SSA}).

\begin{table}[t]
	\caption{Annealing parameters for SSQA}
	\label{tb:param}
	\centering
	\begin{tabular}{c|c}
		\hline
		Parameter & Value \\
		\hline 
		$J_{\perp min}$ & 0 \\
		$J_{\perp max}$ & 0.5 \\
		$\beta$ & 3 \\
		$d$ & 1 \\
		$I_0$ & 2 \\
		$n_{rnd}$ & 1 \\
		\hline
		$\tau$ & 50 to 100 \\
		$R$ & 2 to 50 \\
		EC & 10,000 to  40,000 \\
		\hline
		SC  & $\mathrm{EC}/R$ \\
		Number of cycles per iteration &   $\tau\cdot(\beta+1)$ \\
		Number of  iterations &  $\mathrm{SC}/(\tau\cdot(\beta+1))$\\
		\hline
	\end{tabular}
\end{table}

With the established $h$ and $J$, simulated annealing begins, searching for the global minimum energy. 
In SSQA, the spin states are updated in accordance with equation (\ref{eqn:SC-QMC}) during the annealing process.
The parameters for SSQA are summarized in \cref{tb:param} and are described as follows. 
During each iteration, the scheduling parameter $J_{\perp}$ is incrementally increased from a minimum value, $J_{\perp min}=0$, to a maximum, $J_{\perp max}=0.5$. 
Within the iteration, $J_{\perp}$ is updated as $J_{\perp}(t+1) = J_{\perp}(t) + (J_{\perp max}-J_{\perp min})/\beta$ at every $\tau$ cycle. 
For instance, each iteration involves 400 cycles with $\beta=3$ and $\tau=100$.
When a new iteration begins, $J_{\perp}$ is reset to $J_{\perp min}$. 
Parameters specific to SSQA, such as $d=1$, $I_0=2$ and $n_{rnd}=1$, are determined based on a grid search. 
Other parameters, like $\tau$ and $R$, are varied to evaluate the performance of SSQA.
All simulations are conducted using MATLAB R2020b on an AMD Ryzen-9 5950X at 3.4 GHz with 32 GB of memory.
%

As the number of replicas $R$ increases, the computation cost of SSQA  based on equation (\ref{eqn:SC-QMC}) also increases.
To objectively assess the dependency of performance on $R$, we define the term `Equivalent Cycles (EC)' as follows:
\begin{equation}
	\mathrm{Equivalent \ Cycles \ (EC)} = R \times \mathrm{Simulation \ Cycles \ (SC)}.
	\label{eqn:EC}
\end{equation}
For instance, with EC set to 20,000, SC amount to 2,000 for $R=10$ and drop to 500 for $R=40$. 
Given the same EC value, the computational cost (i.e., simulation time) can be nearly identical for any arbitrary value of $R$.

\subsection{Simulation analysis of SSQA}

\begin{figure}[t]
	\centering
	\includegraphics[width=0.6\linewidth]{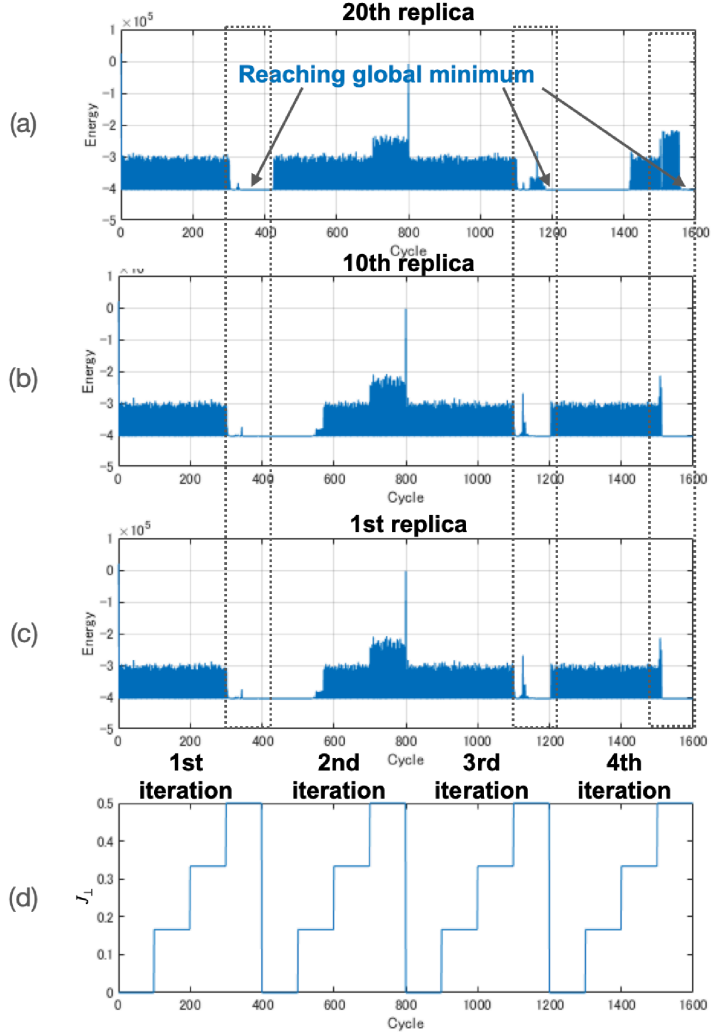}
	\caption{Energy versus cycles  in SSQA for a GI problem of $N=2,500$ with $R=25$: (a) 20th replica, (b) 10th replica, (c) 1st replica, (d) $J_{perp}$.  In each iteration, $J_{\perp}$ is increased with $\tau=100$. At the 1st, 3rd, and 4th iterations, these replicas reach the global minimum energy with different timing.}
	\label{fig:waveform}
\end{figure}

\cref{fig:waveform} displays energy versus cycles for the 1st, 10th, and 20th replicas in SSQA applied to a GI problem of $N=2,500$ with $R=25$.
The simulation encompasses four iterations, each of which consists of 400 cycles with $\tau=100$. 
Each replica strives to minimize its energy as $J_{\perp}$ increases. 
A replica achieving a lower energy level can influence others in their quest to reach the global minimum energy.
When $J_{\perp}$ reaches 0.5, each replica is significantly affected by its neighboring replicas. 
At the 1st, 3rd, and 4th iterations, these replicas reach the global minimum energy at different timings.

\begin{figure}[t]
	\centering
	\includegraphics[width=0.6\linewidth]{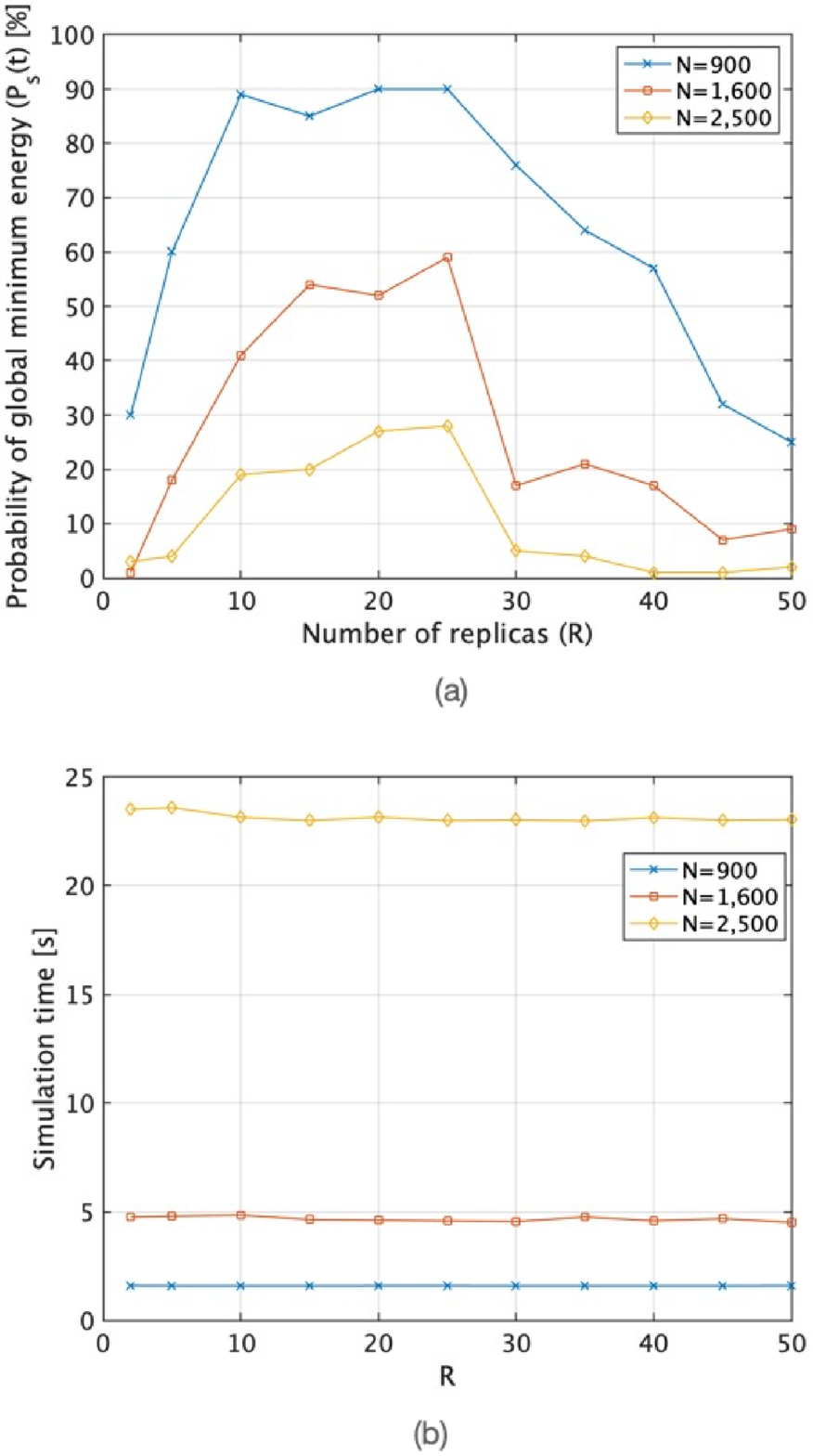}
	\caption{Performance of SSQA in GI problems for $\mathrm{EC}=20,000$ and 100 trials: (a) probability of global minimum energy ($P_a(t)$) versus Number of replicas ($R$) and (b) simulation time vs. $R$. $R=20$ can be a good parameter for this simulation condition.}
	\label{fig:GI_acctime}
\end{figure}

\cref{fig:GI_acctime} (a) displays the probabilities of global minimum energy ($P_s(T)$) versus the number of replicas $R$ in SSQA for EC set at 20,000 and 100 trials with varying problem sizes of $N$. 
As $N$ is the square of the number of nodes in the GI problems, $N=900$ implies that the isomorphism of two 30-node graphs is being checked.
\cref{fig:GI_acctime} (b) shows the simulation time $T$ in SSQA for  EC of 20,000. 
The simulation time remains almost equivalent for all $R$ as assumed in \cref{eqn:EC}.
Considering problem size, $P_s(t)$ decreases and the simulation time increases as $N$ increases.

To evaluate the performance of SSQA,  time-to-solution (TTS) is selected as a performance metric \cite{TTS}.
TTS  is the approximated time to obtain the global minimum energy of the problem Hamiltonian and is defined as follows:
\begin{equation}
	\mathrm{TTS} = t \frac{\mathrm{ln}(1-P_T)}{\mathrm{ln}(1-P_s(t))},
	\label{eqn:TTS}
\end{equation}
where $P_T$ is the probability of finding the global minimum energy at least one time in $T$ trials within the simulation (execution) time $t$.
Using the simulation results of $P_s(t)$ and $t$, TTS is calculated with a target specification of $P_T$.
When $P_s(t)$ is 0\% or 100\%, TTS cannot be calculated.
For example, when $P_s(t)$ is equal to 100\%, TTS approaches 0 as the denominator of TTS is negative infinity.

\begin{figure}[t]
	\centering
	\includegraphics[width=0.6\linewidth]{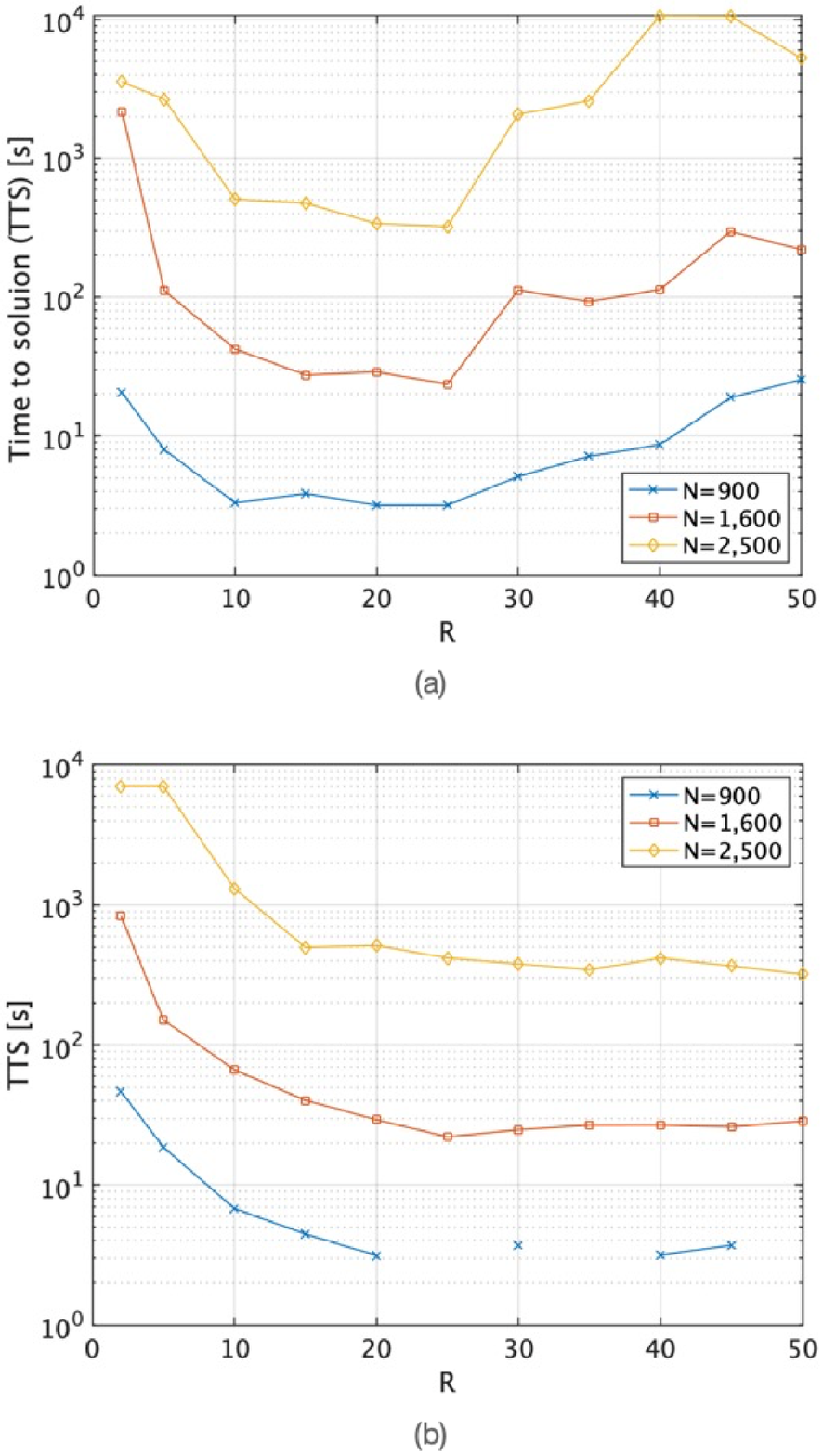}
	\caption{Time to solution (TTS) [s] versus $R$ of SSQA in the GI problems with $P_T=0.99$: (a) $\mathrm{EC}=20,000$ and (b) $\mathrm{EC}=40,000$. Based on the results, $R=25$ can be a good parameter for SSQA.}
	\label{fig:GI_TTS}
\end{figure}

\cref{fig:GI_TTS} displays TTS versus the number of replicas $R$ in SSQA for the GI problems, with $P_T=0.99$. 
When EC  is set to 20,000, TTS is minimized at an $R$ value of approximately 20. 
When EC  is set to 40,000, some TTS values cannot be calculated because $P_s(t)$ equals 1 in the case of $N=900$. 
This implies that an EC of 40,000 is too large for smaller problem sizes. 
An $R$ value of approximately 30 is ideal when EC is set to 40,000. 
Based on these findings, $R=25$ is selected for comparing SSQA with conventional methods, which will be described in the subsequent subsections.

\begin{figure*}[t]
	\centering
	\includegraphics[width=1.0\linewidth]{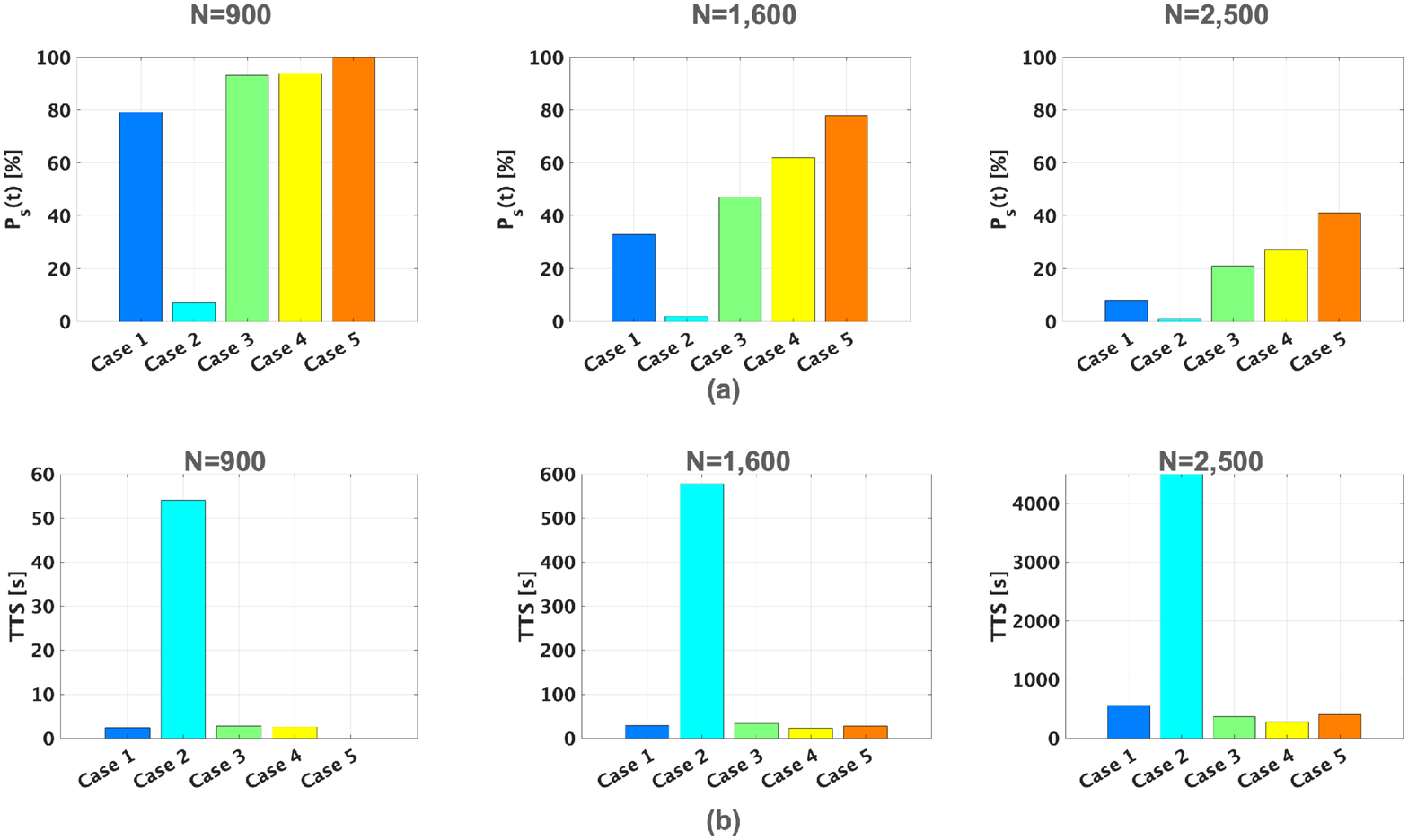}
	\caption{Probability of achieving the global minimum energy $P_s(t)$ and the corresponding TTS using SSQA with $R=25$ for $N=900$, $N=1,600$, and $N=2,500$. Note that case 2, which uses only one iteration, rarely reaches the global minimum energy. Additionally, the optimal  EC and $\tau$ can vary depending on the problem size $N$.}
	\label{fig:SSQA_performance}
\end{figure*}

\begin{table}[t]
	\caption{Five cases of different EC and $\tau$ with $R=25$ used for evaluation in  \cref{fig:SSQA_performance}.}
	\label{tb:case}
	\centering
	\begin{tabular}{c||c|c|c|c|c}
		\hline
		Parameter & Case 1 & Case 2 & Case 3 & Case 4 & Case 5 \\
		\hline
		\hline
		EC & 10,000 & 10,000 & 20,000 & 20,000 & 40,000 \\
		\hline
		$\tau$ & 50 & 100 & 50 & 100 & 100\\
		\hline
		\# iterations &  2 & 1 & 4 & 2 & 4 \\
		\hline
	\end{tabular}
\end{table}

\cref{fig:SSQA_performance} illustrates the probability of achieving the global minimum energy $P_s(t)$ and the corresponding TTS across different problem sizes $N$. 
These simulation results are based on 100 trials for five different cases. 
The parameters for these five cases, all with $R=25$, are summarized in \cref{tb:case}.
The number of iterations is calculated using $(\mathrm{EC}/R/(\tau\cdot(\beta+1)))$ summarized in \cref{tb:param}.
From the simulation results, it appears that case 2 rarely achieves the global minimum energy. 
This is likely due to it having only a single iteration, while all other cases have multiple iterations. 
When comparing case 1 and 3, case 3 attains a higher $P_s(t)$ due to its larger EC.
In terms of TTS, case 1 demonstrates a smaller TTS for $N=900$ and $N=1,600$, suggesting that an EC of 20,000 may be too large for smaller problem sizes. 
When comparing case 3 and 4, case 4 achieves a better $P_s(t)$, even though both cases use the same EC, indicating that a larger $\tau$ might lead to better convergence to the global minimum energy.
Comparing case 4 and 5, case 4 achieves smaller TTS for $N=1,600$ and $N=2,500$, whereas case 5 shows a higher $P_s(t)$. 
Based on these simulation results, the optimal EC and $\tau$ can vary depending on the problem size $N$.

\subsection{Comparisons}

\begin{table*}[t]
	\caption{Performance comparisons of SA \cite{SA1}, QA \cite{QA_GI}, SSA \cite{SSA}, and SSQA in GI problems with $P_T=0.99$.}
	\label{tb:comp}
	\centering
	\begin{tabular}{c||c|c|c|c|c|c|c|c|c|c|c|c}
		\hline
		$N$ & \multicolumn{3}{c|}{SA}  & \multicolumn{3}{c|}{QA} & \multicolumn{3}{c|}{SSA} & \multicolumn{3}{c}{SSQA (proposed)} \\
		\cline{2-13}
		& $P_s(t)$ & $t$  & TTS  & $P_s(t)$  & $t$  & TTS 	& $P_s(t)$ & $t$  & TTS & $P_s(t)$  & $t$  & TTS \\
		&  [\%] & [s] &  [s] &  [\%] & [s] & [s]	&  [\%] & [s] & [s]	& [\%]  &  [s] &  [s] \\
		\hline
		\hline
		25 &  100 & 0.0876 & $\approx$ 0 & 98 &1& 1.17 & 100 & 0.0593 & $\approx$ 0 & 100 & 0.131 & $\approx$ 0 \\
		\hline
		100  & 43 & 0.123 & 1.01& 0 &1& - & 100 & 0.121 &  $\approx$ 0 & 100  & 0.198 & $\approx$ 0 \\
		\hline
		400  & 2 & 0.649 & 148  &  0&1 & - & 100 & 0.717 & $\approx$ 0 & 100 &0.836 & $\approx$ 0 \\
		\hline
		625 & 3 & 1.35 & 203 & N/A & N/A & N/A & 92 & 1.48 & 2.71 & 100 & 1.61 & $\approx$ 0 \\
		\hline
		1225 & 0 & 4.99 & -  & N/A & N/A & N/A & 49 & 5.30 & 36.2 & 95 & 5.42 & 8.33\\
		\hline
		2025 & 0 & 19.4 & - & N/A & N/A & N/A & 6 & 22.7 & 1690 & 51 & 22.7 & 146 \\
		\hline
		2500 & 0 & 36.1 & - & N/A & N/A & N/A & 0 & 45.6 & - &  41 & 46.4 &  405 \\
		\hline
	\end{tabular}
\end{table*}

The proposed SSQA method is compared with traditional SA \cite{SA1}, QA \cite{QA_GI}, and SSA \cite{SSA} for solving GI problems.
To ensure a fair comparison, the SA and SSA methods are simulated on the same computer that is used for SSQA.
In SA, the temperature parameter $T_0(t)$ is gradually decreased by $\Delta_{IT}$ according to the schedule $T_0(t+1) = 1/(1/T_0(t)+\Delta_{IT})$ at each cycle.
During each of these cycles, the algorithm randomly flips a spin state and accepts the new state if the new energy ($E_{new}$) is lower than the current energy ($E_{cur}$), or with a probability of $\mathrm{exp}(-(E_{new}-E_{cur})/T_0(t))$ if it is higher.
The SA process starts with an initial temperature of 1,000 and ends with a final temperature of 0.1.
On the other hand, in SSA, a pseudoinverse temperature parameter $I_0(t)$ is gradually increased for each iteration.
$I_0$ follows the update rule $I_0(t+1) = (1/\beta)\cdot I_0(t)$ every $\tau$ cycles, ranging from $I_{0min}=1$ to $I_{0max}=16$ with parameters $\tau=10$ and $\beta=0.5$.
For SSA, this means that each iteration consists of 50 cycles.
Another parameter, specific to SSA, is $n_{rnd} = 1$, as suggested by \cite{SSA}.
For QA, we refer to and compare with the experimental results presented in \cite{QA_GI}.

\cref{tb:comp} summarizes the performance of the different methods – SA, QA, SSA, and SSQA – in solving GI problems.
In the case of QA, a 504-qubit D-Wave Two machine was used, and an execution time $t$ of 1 second was allocated.
Both SA and SSA are simulated for a total of $40,000$ cycles.
SSQA, in contrast, is configured to run for $\mathrm{SC}=1,600$ and $R=25$, which effectively corresponds to $\mathrm{EC}=40,000$, as explained in \cref{eqn:EC}.
For SA, SSA, and SSQA, $P_s(t)$ values are obtained through an average of 100 trials.

When comparing SSA and SSQA, the traditional SA method requires less computational cost as it employs a serial spin-update process.
$P_s(t)$ significantly decreases as the problem size $N$ increases, leading to an increase in TTS.
The selected 40,000 simulation cycles appear to be insufficient for the SA method to converge for large-scale problems.
In QA, the D-Wave machine can process approximately 500 spins (bits) for combinatorial optimization problems, but it only supports neighborhood connections between spins, as reported in \cite{QA_GI}.
Given that the GI problems in this study are modeled using fully connected spins, adjustments to the model are necessary for it to be compatible with the QA machine.
As such, the QA method was only applied to GI problems with 25 spins.
Interestingly, a new D-Wave machine, known as D-Wave 5000, has been released and is capable of handling 5,000 spins \cite{D-Wave5000}.
However, as of now, there are no available reports on the performance of the new machine concerning GI problems, and the issue with fully connected spins might still persist due to hardware limitations.

In contrast, SSA is able to solve considerably larger problems than SA and QA, and it also exhibits a smaller TTS compared to SSQA for small-scale problems.
However, as the problem size $N$ increases, SSQA outperforms SSA by achieving higher $P_s(t)$ values and smaller TTS.
For instance, with $N=2,025$, SSQA reduces the TTS by a remarkable 91.4\% compared to SSA.
The difference in TTS between SSA and SSQA continues to widen as $N$ grows.
Consequently, the proposed SSQA algorithm demonstrates to be significantly more effective than SSA when tackling large-scale combinatorial optimization problems.
Compared with SA, SSQA can solve 25-times larger nodes  with similar $P_s(t)$.
Additionally, SSQA can handle problems with two orders of magnitude larger number of spins compared to QA, making it an extremely powerful tool for solving large and complex GI problems.

\section{Conclusion}
\label{sec:conc}

In this paper, we present SSQA, a new method devised for tackling large-scale combinatorial optimization problems.
SSQA employs an innovative spin-state update algorithm rooted in integral stochastic computing, which uses randomization and approximation, making it adept at solving complex optimization problems.
We have experimentally  evaluated the impact of varying the number of replicas on the performance through simulations. 
The findings suggest that using 25 replicas is particularly effective for GI problems, a common class of combinatorial optimization problems.
Additionally, SSQA is compared to conventional SSA, QA using a 504-qubit D-Wave Two machine, and traditional SA. 
SSQA exhibits an order of magnitude smaller TTS than SSA and can solve problems with roughly 100-times more nodes compared to QA and 25-times more nodes than SA.

Looking ahead, the development of a large-scale hardware implementation of SSQA could be promising, as it holds the potential to be a fast and efficient solver for real-world combinatorial optimization challenges.

\section*{Acknowledgment}

This work was supported in part by JST CREST Grant Number JPMJCR19K3, and JSPS KAKENHI Grant Number JP21H03404.

\bibliographystyle{unsrtnat}
\bibliography{SSQA}

\end{document}